# Source Illusion Devices for Flexural Lamb Waves using Elastic Metasurfaces

Yongquan Liu[1], Zixian Liang[2], Fu Liu[1], Owen Diba[1], Alistair Lamb[1], Jensen Li[1*]

[1] School of Physics and Astronomy, University of Birmingham, Birmingham, United Kingdom, B15 2TT
[2] College of Electronic Science and Technology, Shenzhen University, Shenzhen, China, 518060
[*] Email: j.li@bham.ac.uk

Metamaterials with the transformation method has greatly promoted the development in achieving invisibility and illusion for various classical waves. However, the requirement of tailor-made bulk materials with extreme constitutive parameters associated to illusion designs hampers its further progress. Inspired by recent demonstrations of metasurfaces in achieving reduced versions of electromagnetic cloaks, we propose and experimentally demonstrate source illusion devices to manipulate flexural waves using metasurfaces. The approach is particularly useful for elastic waves due to the lack of form-invariance in usual transformation methods. We demonstrate metasurfaces for shifting, transforming and splitting a point source with "space-coiling" structures. The effects are found to be broadband and robust against a change of source position, with agreement from numerical simulations and Huygens-Fresnel theory. The proposed approach provides an avenue to generically manipulate guided elastic waves in solids, and is potentially useful for applications such as non-destructive testing, enhanced sensing and imaging.

The ability to make objects invisible has captured our imagination since ancient times, and becomes reachable very recently due to the appearance of metamaterials[1] together with an emerging design framework of transformation optics[2-5]. A family of more general illusion effects has also been established based on the transformation approach. Either scattering[6,7] or source radiation[8] can be transformed with the help of coordinate mappings to prescribed target patterns. Furthermore, the transformation approach has been extended to control acoustic waves[9-11] and heat flows[12-14]. Despite these huge successes, the transformation approach cannot work automatically for some classical waves, with the elastic waves being a well-known example. Although some elastic cloaks have been studied and designed with approximations[15,16], the transformation method, together with the technique of elastic metamaterials[17-22], is non-universal to manipulate elastic waves. Because the governing equation in elastodynamics (i.e., the Navier's equation) cannot keep its form invariant under a general coordinate transformation[23], unless further approximations[24] are applied or the proposed Willis media [25] can be realized and employed with complex cross-coupling response. It makes the application of transformation approach on elastic waves non-trivial and more generic illusion applications, which may use negative-index metamaterials with unavoidable loss and fabrication issues, very challenging.

On the other hand, the recently proposed metasurface[26-33], a kind of artificial sheet material with subwavelength-scaled patterns and thickness, can modulate wavefronts on demand through specific boundary conditions. The metasurface approach shows the simplicity in fabrication with low-loss, compact form-factor, but without losing the extreme functionality of bulk metamaterials. It has already been demonstrated as an effective way to manipulate wavefront in both optics and acoustics, including generating anomalous refraction or reflection, arbitrary surface plasmon profiles, high efficiency holograms[26-32] and even an ultrathin invisibility cloak[33,34], which is a much simplified version of a carpet cloak originally using bulk metamaterials in transformation approach[35]. Therefore, the extension of metasurfaces to elastic waves is expected to be very effective to demonstrate non-trivial wave phenomena in elastic waves, which associate to richer physics [20-22] due to its more degrees of freedom than acoustic and electromagnetic counterparts and a wide range of potential applications in non-destructive testing, enhanced sensing and imaging[18, 36-40]. Until very recently, Zhu and Semperlotti[41] have presented the first experimental demonstration of elastic metasurfaces to achieve anomalous refraction of guided waves in solids, based on a cross-"polarization" conversion (from symmetric $S_0$ to antisymmetric $A_0$ mode) with a transmission efficiency around 18%, comparing to the theoretical maximum efficiency 25% in the case of electromagnetic waves[42,43]. A higher transmission efficiency, together with a suitable design framework, will be an enabling key to a much wider range of applications of elastic metasurfaces.

In this article, we develop a theoretical framework and experimental realization of source illusion devices using elastic metasurfaces. Our design is based on the "space coiling" mechanism[44,28-30], originally developed for air-borne acoustic waves, for designing microstructures without employing polarization conversion. It allows high transmission efficiency and large working frequency bandwidth. By incorporating an additional Fabry-Pérot resonance, we can further modulate transmission phase and amplitude independently. This flexible control, together with the metasurface approach, requires neither the fore-mentioned form-invariance of elastodynamics nor bulk metamaterials with negative indices, yet allows us to demonstrate a series of all-angle elastic illusion effects, including source shifter, transformer and splitter. Despite designed for particular source at specific frequencies, these devices work reasonably well subject to a deviation from the designed configuration when there is a change in the working frequency or a change in the position of the physical source.

**Illusion devices using space-coiling metasurfaces.** Fig. 1(a) shows a schematic diagram illustrating the design strategy of illusion devices using metasurfaces. The target wave profile, living in the virtual space, is described by its phase profile $\phi_{tar}(x, y)$ (left hand side of Fig. 1(a)). At the moment, we neglect the contribution of amplitude profile for simplicity. Suppose that we can generate an incident wave field characterized by its phase $\phi_{in}(x, y)$, a ring-type metasurface with the phase discontinuity of $\Delta\phi = \phi_{tar} - \phi_{in}$, being evaluated at the location of metasurface, will turn the incident field into an arbitrary target profile (right hand side of Fig. 1(a)). By doing this, an observer outside the metasurface (the black dashed circle) finds the same target field pattern in the physical space, as designed in the virtual space. Next, the phase discontinuity can be achieved by a particular metasurface design, which should cover an entire phase change of $2\pi$ with preferably unit transmission amplitude. Previous design[41] utilized resonant unit cells with mode conversion from $S_0$ mode to $A_0$ mode to achieve a full coverage of phase change from 0 to $2\pi$, which leads to low power of transmitted waves. In light of this, we design elastic metasurfaces with negligible mode-conversion based on the space coiling mechanism[44]. Fig. 1(a) also shows the schematic of our design to realize the phase discontinuity. Each unit cell of the ring-type metasurface consists of a curved thin bar with identical width $d$ and height of turns $h$. It modulates the propagation phase of the guided asymmetric Lamb wave ($A_0$ mode) within a 3mm-thick acrylic plate. To illustrate its working mechanism, Fig. 1(b) shows simulations of the out-of-plane displacement field, which corresponds to the $A_0$ antisymmetric mode Lamb wave, produced by different heights $h$ of three-turn unit cells. The wave is restricted to propagate along a zigzag path shaped by the unit cell. By increasing the height of turns $h$, we can increase the phase discontinuity $\Delta\phi$ of transmitted waves due to the increase of path length. In consequence, the crests of the wave profiles (dashed black lines) in the transmission side gradually shift to the left. Fig. 1(c) describes the phase discontinuity $\Delta\phi$ (black lines) and the amplitude of transmitted $A_0$ waves $|t|$ (red lines) of unit cells versus $h$ with chosen geometric parameters of $d = 1$mm, $l = 15$mm and $H = 5.89$mm at 8 kHz and 9 kHz. It is worth to note that the average value of $|t|$ is over 0.85 with respect to the incident $A_0$ wave, corresponding to a transmission power above 0.72. It infers that this unit cell can redirect the wave front without significant reflected waves. It enables the illusion effects demonstrated in this work. Moreover, the phase discontinuity $\Delta\phi$ can keep its monotonic trend in $h$ and a full coverage of $2\pi$ with a variation of frequency.

**Source shifter.** As the first example, an illusion device for shifting the position of a wave source virtually, termed as a source shifter, is constructed, with its effect schematically shown in Fig. 2(a). Previous illusion optical devices[8] are composed of bulk materials with inhomogeneous parameters, resulting from the transformation approach. Here we achieve the illusion effect using a ring-shape metasurface to shift the source horizontally by a distance $\Delta S$ to the left. Suppose, we take a special case where a physical source at origin $(0,0)$ is shifted to a virtual one at $(-\Delta S, 0)$. The metasurface, of radius $R$, should possess phase discontinuity $(\Delta \phi)$ as a function of azimuthal angle $\theta$, given by

$$\Delta\phi(\theta) = \frac{2\pi}{\lambda}(\sqrt{(R\cos\theta + \Delta S)^2 + (R\sin\theta)^2} - R) \qquad (1)$$

where $\lambda$ is the working wavelength in the acrylic plate. Here, we set $R =57$ mm, $\Delta S =25$ mm and $\lambda =26.0$ mm (corresponding to 12 kHz). Then the metasurface with 48 unit cells possessing the required phase change is designed and fabricated. Detailed geometric parameters of the unit cells (with photo) are provided in Supplementary Fig. 2. The out-of-plane velocity field is measured by a laser scanning vibrometer (Polytec PSV-400), and is shown in Fig. 2(b). For comparison, crests of the target pattern (dashed gray lines) are also plotted. Inside the ring, it is the circular wavefront expected from the point source before transformation. Circular wave fronts are also measured outside metasurface. They are all centered at about 25.2 mm on the left of the exciting point source, which is very close to the target value 25 mm. The ripples can be further improved by using unit cells of a smaller lattice constant with respect to the wavelength.

To illustrate the robustness of our metasurface approach, Fig. 2(c) shows the normalized wave field at another frequency of 10 kHz. Apart from the change of wavelength from 26.0mm to 28.7mm, similar wave pattern, wavefronts with common center, are still found to match the target one (presented as dashed gray lines), with an illusion about shifting the point source horizontally to the left by 27.5mm. More interestingly, if we place the source at some other locations, e.g. $(x_0,0)$, the wave fronts outside are still found to have the shifted common center at around $(x_0 - \Delta S, 0)$. It is found to work well from $x_0 = $ -10 mm to 30 mm. Fig. 2(d) presents the measured wave pattern for a particular case $x_0 = $ 25mm, where the common center of wave fronts is found at the origin. Based on the Huygens-Fresnel diffraction theory (see Supplementary Note 1 for details), the wave profile outside the metasurface can be calculated, with the crests plotted as dashed gray lines in Fig. 2(d), coinciding those of the experimental result. The associated numerical simulations of the experimental configurations in Fig. 2 are provided in more details in Supplementary Fig. 3.

**Source transformer.** Our elastic metasurfaces can also be used to transform a point source to a prescribed target wavefront. Here we impart an additional angular momentum[26] to the point source as an intuitive example. The observer outside perceives it is a point source with spiral wave front. In this case, the phase discontinuity imparted by the metasurface should be simply $\Delta\phi(\theta) = L\theta + \phi_0$ with $L$ being the additional angular momentum (integer) and $\phi_0$ being an arbitrary constant. We set $L = 6$, $R = 60$ mm and the operating frequency at 8 kHz, the metasurface with 48 unit cells can be designed and fabricated with a similar procedure (with details in Supplementary Fig. 4). Fig. 3(a) shows the numerical simulations of the $A_0$ mode wave field excited by a point source at the center, and Fig. 3(b) illustrates the corresponding experimental results. In the external region of the metasurface, the cylindrical wave shapes into a spiral with evenly-distributed six branches. All branches are centered at the original point. It is noted that the number of branches equals the additional momentum $L$, equivalently the wave undergoes a phase change of $2\pi L$ for walking around one cycle in the azimuthal direction. For comparison, we also plot the crests of the theoretical pattern as dashed lines in Figs. 3(a) and (b), which coincides with both the numerical and experimental results.

The designed metasurface also works with a variation of the working frequency and source position. For instance, we change the experimental frequency to 10 kHz for the same designed metasurface (Fig. 3(c)), we can still achieve a similar six-branched spiral wave field, except the wavelength (the spacing between crests) is changed from 32.4mm to 28.7mm. In another case, we place the source at (-15, 0) mm instead of the origin, the measured wave pattern is shown in Fig. 3(d). Again, a spiral wavefront with six branches, indicating $L = 6$, is found. The wave field with a generic source position can actually be explained by the Huygens-Fresnel principle, with the phase outside the metasurface being approximated as

$$\phi_{\text{out}}(x,y) = \Delta\phi + k(\sqrt{x^2 + y^2} - R + \sqrt{(R\cos\theta - x_0)^2 + (R\sin\theta - y_0)^2}) \quad (2)$$

where $(x_0, y_0)$ is the position of the actual point source, $\Delta\phi$ and $R$ are the phase discontinuity and the radius of the ring-type metasurface. Crests of the theoretical profile predicted by Eq. (2) are plotted as dashed gray lines in Fig. 3(d), matching the experimental results.

**Source splitter.** It should be noted that the two demonstrated metasurfaces make illusion as a single point source as target. In both cases, the amplitude of the wave pattern on the metasurface boundary only changes slightly. Hence, as a prerequisite of keeping a high value of the transmittance, we just need to tune the phases of metasurfaces. As a result, we only need to utilize three-turned unit cells for the design of the above two metasurfaces. As a further step, we now consider the case with changing wave amplitude on the metasurface boundary as our final example. As shown in Fig. 4(a), a source splitter is designed to mimic a wave pattern of two sources with opposite phases from a single physical point source. The two virtual sources are set at positions (0, 20)mm and (0, -20)mm respectively, and the designed frequency is 8 kHz. Due to the interference of the two target virtual sources, both the amplitude and phase vary greatly on the metasurface boundary. Thus we need to have unit cells possessing adequate phase shift (0 to $2\pi$) and transmittance (0 to 1) simultaneously. By changing the width $d$ and the number of turns of the curved bar to incorporate additional Fabry-Pérot resonance, very low to high transmission amplitude can be achieved. As an example, Fig. 4(b) shows the phase discontinuity $\Delta\phi$ (black line) and amplitude of the transmitted wave $|t|$ (dashed red line) of a two-turn unit cell with $l = 15$mm and $d = 1$mm, as a function of $h$ at 8 kHz. The amplitude of transmitted waves can be as low as zero when $h = 3.76$ mm. Based on this mechanism, six types of unit cells with different width $d$ and different number of turns are utilized to construct this metasurface. Details of each unit cell are provided in Supplementary Fig. 6. Fig. 4(c) presents the experimentally measured $A_0$ wave field at 8 kHz. The wave is excited by a point source at the origin, and an interference pattern is observed outside the metasurface. Because of the low transmittance of some unit cells, the amplitude inside the metasurface is much larger than that outside. The theoretical wave fronts at zero phase are plotted as dashed lines outside the metasurface for comparison. It is worth to note that the amplitude on x-axis is nearly zero due to the use of low-transmission unit cells, such that waves propagating in the upper and lower plane can has a $\pi$ phase lag with each other. Although resonance is employed, there is still an acceptable bandwidth for operation, e.g. at 8.2 kHz (Fig. 4(d)), except the wavelength of the circular wave front is changed from 32.4 mm to 31.9 mm.

The illusion devices described above are just a few examples of the proposed strategy to make illusions using elastic metasurfaces. More elaborated illusion effects become possible and are expected in future works through our framework, such as combining the power of two or more sources for omnidirectional radiation, or making an arbitrary scatterer appearing as a very different one originally in the electromagnetic and acoustic cases [6, 8]. It is worth to note that bulk metamaterials with extreme or negative refractive indices are often required for illusion devices from the transformation approach and are avoided here. Nevertheless, the metasurfaces proposed in this paper are quite compact, with thickness less than half of the wavelength at the operation frequency.

For the cases of source shifter and source transformer, space-coiling metasurfaces are designed without using the resonant mechanism. Thus they are basically broadband and display high power output to redirect the wave fronts smoothly. Actually, we have found that the source shifter experimentally works from 10 kHz to 13 kHz with an error less than 10 % in the shifting center of wavefront. The six-branched spiral pattern is experimentally observed in the source transformer over a frequency range of 6 kHz to 10.5 kHz, a bandwidth more than 50% of the central frequency. Moreover, when we change the position of the source, the corresponding target phase $\phi_{tar}$ and the incident phase $\phi_{in}$ are found to have approximately the same difference evaluated on the metasurface so that the same devices are still valid around the initially designed position of the source, as illustrated in Figs. 2(d) and Fig. 3(d), as generic illusion effects. In fact, the position of the source can be arbitrarily set around the center. For example, as shown numerically in Supplementary Fig. 5(b), the source transformer can still preserve a six-branched spiral pattern when setting the source at a position of (10, 10)mm.

In summary, we have developed a generic framework to make source illusions for $A_0$ mode Lamb waves using compact elastic metasurfaces. As examples, we designed and experimentally demonstrated three all-direction illusion devices to shift the position of a point source, to impart additional angular momentum, and to split a point source into two, respectively. The demonstrated illusion effects are robust against a change of working frequency and position of source. Without requiring form-invariance of the governing equation in elastodynamics to achieve illusion effects, the proposed approach paves the way to "transformation approach"-type applications to seek more functionalities of elastic metasurfaces. Our investigations may find potential applications in non-destructive testing, enhanced sensing and imaging.

**Methods**

**Numerical simulations.** All the simulations performed in the paper are obtained using COMSOL Multiphysics. Throughout the paper, we study 3 mm elastic plates built out of a kind of 3D printed material (VeroBlue RGD840), whose properties are measured as: Young's modulus E= 2.72 GPa, Poisson's ratio ν=0.38 and density ρ=1190 kg/m$^3$. At frequencies of 8 kHz and 12 kHz, the wavelengths of the $A_0$ mode Lamb wave are 32.4 mm and 26.0 mm, respectively. For the simulations to calculate the phase shift and the transmitted amplitude of unit cells, perfect matched layers (PML) are used on both end of every beam-like model, and periodic boundary condition at the top and bottom boundaries are applied. For all simulations on the wave patterns of metasurfaces, PML are used on all outer boundaries.

**Fabrication and experiemental set up.** The experimental samples are all 280×180×3 mm$^3$ flat plates and printed directly using a Stratasys Objet30 Pro 3D printer. A Polytec PSV-400 laser scanning vibrometer is used to generate a 20-cycle tone burst $w(t) = A_0[1 - cos(\frac{2\pi f_c}{20}t)]sin(2\pi f_c t)$, where $f_c$ is the central frequency. Through the amplification of a power amplifier (KH 7602M), the signal transfers to the 12mm Piezo Elements Sounder Sensor bonded on the back side of the sample. To minimize reflected waves from the outer boundary of sample, each edge of sample is covered by a layer of blue-tack. The wave profiles are measured using a spatial resolution of approximately 2.5 mm (corresponding to about 13 and 10 points per flexural wavelength at 8 kHz and 12 kHz, respectively). The sampling frequency in the time domain was set at 512 kHz, while an ensemble average with 20 samples was used at every scanning point. By transforming the measured wave profile from the time to the frequency domain, the normalized amplitude of the measured field (240×160 mm$^2$ in the central area, except for the metasurface region) can be obtained.

**References**

1. Smith, D. R., Pendry, J. B. & Wiltshire, M. C. Metamaterials and negative refractive index. *Science* **305**, 788-792 (2004).

2. Leonhardt, U. Optical conformal mapping. *Science* **312**, 1777-1780 (2006).

3. Pendry, J. B., Schurig, D. & Smith, D. R. Controlling electromagnetic fields. *Science* **312**, 1780-1782 (2006).

4. Schurig, D. *et al.* Metamaterial electromagnetic cloak at microwave frequencies. *Science* **314**, 977-980 (2006).

5  Chen, H., Chan, C. T. & Sheng, P. Transformation optics and metamaterials. *Nat. Mater.* **9**, 387-396 (2010).

6. Lai, Y. *et al.* Illusion optics: the optical transformation of an object into another object. *Phys. Rev. Lett.* **102**, 253902 (2009).

7. Li, C. *et al.* Experimental realization of a circuit-based broadband illusion-optics analogue. *Phys. Rev. Lett.* **105**, 233906 (2010).

8  Chen, H., Xu, Y., Li, H. & Tyc, T. Playing the tricks of numbers of light sources. *New J. Phys.* **15**, 093034 (2013).

9. Zhang, S., Xia, C. & Fang, N. Broadband acoustic cloak for ultrasound waves. *Phys. Rev. Lett.* **106**, 024301 (2011).

10  Popa, B. I., Zigoneanu, L. & Cummer, S. A. Experimental acoustic ground cloak in air. *Phys. Rev. Lett.* **106**, 253901 (2011).

11. Sanchis, L. et al. Three-dimensional axisymmetric cloak based on the cancellation of acoustic scattering from a sphere. *Phys. Rev. Lett.* **110**, 124301 (2013).

12. Schittny, R., Kadic, M., Guenneau, S. & Wegener, M. Experiments on transformation thermodynamics: molding the flow of heat. *Phys. Rev. Lett.* **110**, 195901 (2013).

13. Xu, H., Shi, X., Gao, F., Sun, H. & Zhang, B. Ultrathin three-dimensional thermal cloak. *Phys. Rev. Lett.* **112**, 054301 (2014).

14. Han, T., Bai, X., Gao, D., Thong, J. T., Li, B. & Qiu, C. W. Experimental demonstration of a bilayer thermal cloak. *Phys. Rev. Lett.* **112**, 054302 (2014).

15  Farhat, M., Guenneau, S. & Enoch, S. Ultrabroadband elastic cloaking in thin plates. *Phys. Rev. Lett.* **103**, 024301 (2009).

16. Stenger, N., Wilhelm, M. & Wegener, M. Experiments on elastic cloaking in thin plates. *Phys. Rev. Lett.* **108**, 014301 (2012).

17  Liu, Z., Zhang, X., Mao, Y., Zhu, Y. Y., Yang, Z., Chan, C. T. & Sheng, P. Locally resonant sonic materials. *Science* **289**, 1734-1736 (2000).

18. Zhu, R., Liu, X. N., Hu, G. K., Sun, C. T. & Huang, G. L. Negative refraction of elastic waves at the deep-subwavelength scale in a single-phase metamaterial. *Nat. Commun.* **5**, 5510 (2014).

19  Lu, J., Qiu, C., Ke, M. & Liu, Z. Valley vortex states in sonic crystals. *Phys. Rev. Lett.* **116**, 093901 (2016).

20  Yu, S.Y. et al. Surface phononic graphene. *Nat. Mater.* **15**, 1243-1247 (2016).

21  Oh, J. H., Kwon, Y. E., Lee, H. J. & Kim, Y. Y. Elastic metamaterials for independent realization of negativity in density and stiffness. *Sci. Rep.* 6, 23630 (2016).

22  Ma, G., Fu, C., Wang, G., Hougne, P., Christensen, J., Lai Y. & Sheng, P. Polarization bandgaps and fluid-like elasticity in fully solid elastic metamaterials. *Nat. Commun.* **7**, 13536 (2016).

23. Milton, G. W., Briane, M. & Willis, J. R. On cloaking for elasticity and physical equations with a transformation invariant form. *New J. Phys.* **8**, 248 (2006).

24  Hu, J., Chang, Z. & Hu, G. Approximate method for controlling solid elastic waves by transformation media. *Phys. Rev. B* **84**, 201101 (2011).

25  Milton, G. W. & Willis, J. R. On modifications of Newton's second law and linear continuum elastodynamics. *Proc. R. Soc. A* **463**, 855-880 (2007).

26. Yu, N. *et al.* Light propagation with phase discontinuities: generalized laws of reflection and refraction. *Science* **334**, 333-337 (2011).

27  Ni, X., Emani, N. K., Kildishev, A. V., Boltasseva, A., & Shalaev, V. M. Broadband light bending with plasmonic nanoantennas. *Science* **335**, 427-427. (2012).

28. Cheng, Y., Zhou, C., Yuan, B. G., Wu, D. J., Wei, Q. & Liu, X. J. Ultra-sparse metasurface for high reflection of low-frequency sound based on artificial Mie resonances. *Nat. Mater.* **14**, 1013-1019 (2015).

29. Xie, Y., Wang, W., Chen, H., Konneker, A., Popa, B. I. & Cummer, S. A. Wavefront modulation and subwavelength diffractive acoustics with an acoustic metasurface. *Nat. Commun.* **5**, 5553 (2014).

30. Li, Y. et al. Experimental realization of full control of reflected waves with subwavelength acoustic metasurfaces. *Phys. Rev. Appl.* **2**, 064002 (2014).

31. Sun, S., He, Q., Xiao, S., Xu, Q., Li, X. & Zhou, L. Gradient-index meta-surfaces as a bridge linking propagating waves and surface waves. *Nat. Mater.* **11**, 426-431 (2012).

32. Zheng, G., Mühlenbernd, H., Kenney, M., Li, G., Zentgraf, T. & Zhang, S. Metasurface holograms reaching 80% efficiency. *Nat. Nanotechnol.* **10**, 308-312 (2015).

33. Ni, X., Wong, Z. J., Mrejen, M., Wang, Y. & Zhang, X. An ultrathin invisibility skin cloak for visible light. *Science* **349**, 6254 (2015).

34. Orazbayev, B., Mohammadi Estakhri, N., Alù, A. & Beruete, M. Experimental demonstration of metasurface‑based ultrathin carpet cloaks for millimeter waves. *Adv. Opt. Mater.* (2016).

35. Li, J. & Pendry, J. B. Hiding under the carpet: a new strategy for cloaking. *Phys. Rev. Lett.* **101**, 203901 (2008).

36. Ng, C. T. & Veidt, M. A Lamb-wave-based technique for damage detection in composite laminates. *Smart Mater. Struct.* **18**, 074006 (2009).

37. Sukhovich, A., Merheb, B., Muralidharan, K., Vasseur, J. O., Pennec, Y., Deymier, P. A., & Page, J. H. Experimental and theoretical evidence for subwavelength imaging in phononic crystals. *Phys. Rev. Lett.* **102**, 154301 (2009).

38  Mosk, A. P., Lagendijk, A., Lerosey, G. & Fink, M. Controlling waves in space and time for imaging and focusing in complex media. *Nat. Photonics* **6**, 283-292 (2012).

39. Li, J., Fok, L., Yin, X., Bartal, G. & Zhang, X. Experimental demonstration of an acoustic magnifying hyperlens. *Nat. Mater.* **8**, 931-934 (2009).

40  Zhu, J. et al. A holey-structured metamaterial for acoustic deep-subwavelength imaging. *Nat. Phys.* **7**, 52-55. (2011).

41. Zhu, H. & Semperlotti, F. Anomalous refraction of acoustic guided waves in solids with geometrically tapered metasurfaces. *Phys. Rev. Lett.* **117**, 034302 (2016).

42. Monticone, F., Estakhri, N. M. & Alù, A. Full control of nanoscale optical transmission with a composite metascreen. *Phys. Rev. Lett.* **110**, 203903 (2013).

43. Ding X. et al. Ultrathin Pancharatnam−Berry Metasurface with Maximal Cross-Polarization Efficiency. *Adv. Mater.* **27**, 1195-1200 (2015).

44. Liang, Z. & Li, J. Extreme acoustic metamaterial by coiling up space. *Phys. Rev. Lett.* **108**, 114301 (2012).

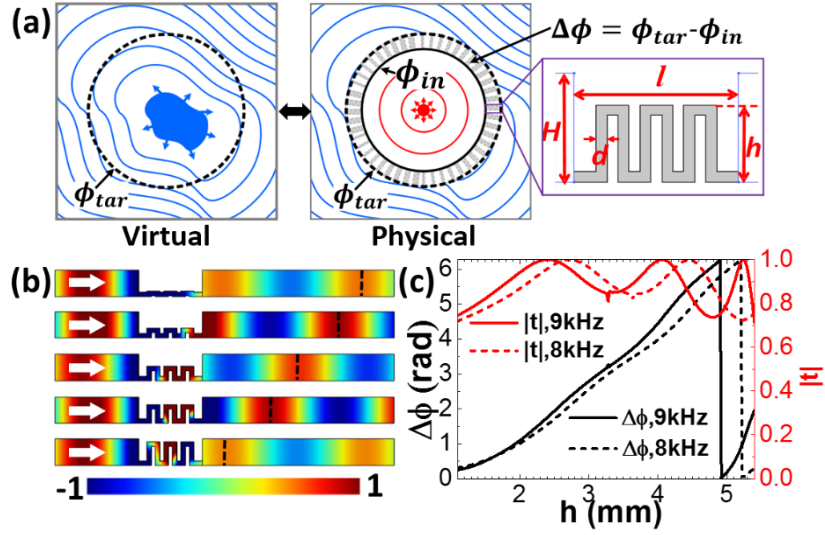

**Figure 1. Space-coiling metasurfaces to achieve illusions.** (**a**) Schematic of making illusions using metasurfaces. The target profile is characterized by its phase $\phi_{tar}(x, y)$, and achieved by the combination of the incident phase $\phi_{in}(x, y)$ and the phase discontinuity $\Delta\phi$ at metasurface. Each unit cell of the metasurface is made of a curved thin bar, zoomed and dimensioned in the right part of the figure. (**b**) Finite element method (FEM) simulations of out-of-plane displacement field produced by different height $h$ of curved unit cells with three turns. The crests of transmitted waves (black dashed lines) shift left with the increase of $h$, which infers the increase of phase discontinuity $\Delta\phi$. (**c**) The phase discontinuity $\Delta\phi$ (black lines) and the amplitude of transmitted waves $|t|$ (red lines) as a function of the height of the curve unit cell $h$. Dashed and solid lines correspond to the results at 8 kHz and 9 kHz, respectively.

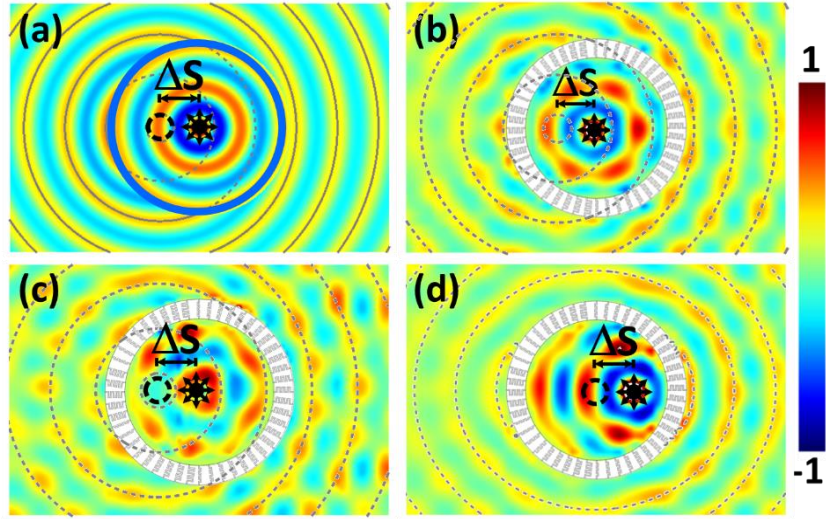

**Figure 2. Source shifter to shift the position of a wave source virtually.** **(a)** Expected analytic profile of the shifter at 12 kHz. The physical point source at the center (black circular point with arrows), is shifted horizontally to the left by $\Delta S = 25$ mm. Crests of the target pattern are plotted as gray lines. **(b)** The normalized amplitude of the measured $A_0$ wave field at the frequency of 12 kHz. The wave source is placed at the center. **(c)** Same as (b), but at the frequency of 10 kHz. **(d)** Same as (b), but the source is placed at (25, 0) mm. In (b)-(d), crests of the theoretical patterns based on the Huygens-Fresnel diffraction theory are plotted as dashed gray lines accordingly.

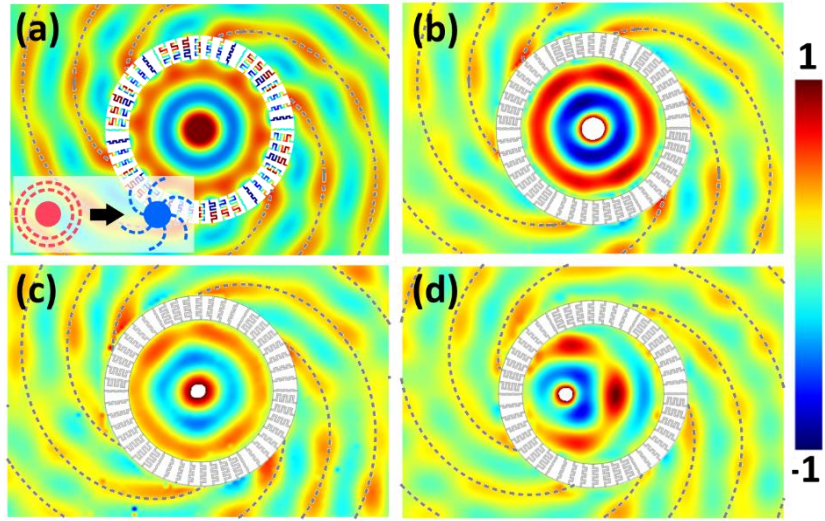

**Figure 3. Source transformer imparting an additional angular momentum to a point source.** **(a)** The simulation results of an illusion metasurface impart an angular momentum of $L = 6$, with the illusion effect schematically shown in the inset. **(b)** The normalized amplitude of the measured $A_0$ wave field at the frequency of 8 kHz. The point wave source is placed at the center. **(c)** Same as (b), but at the frequency of 10 kHz. **(d)** Same as (b), but the source is placed at (-15, 0) mm. In all subplots, crests of wave profiles, from Huygen-Fresnel theory, are plotted as dashed gray lines accordingly.

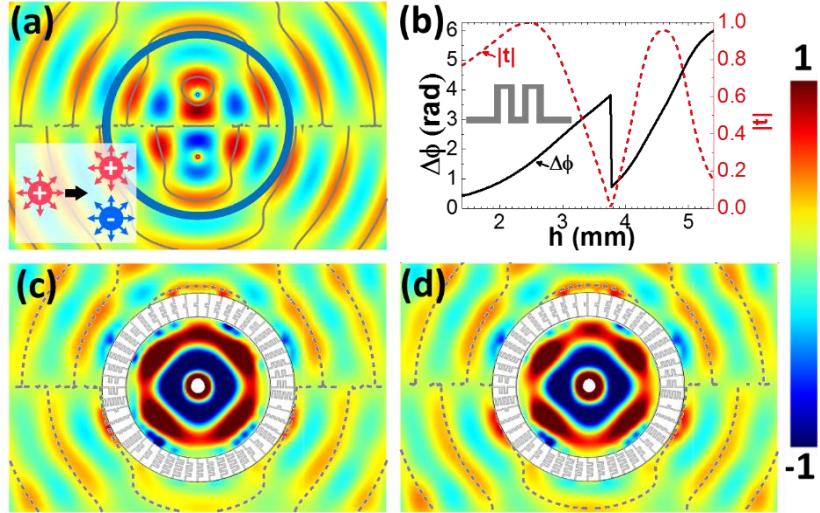

**Figure 4. Source splitter metasurface.** (a) The target profile from two point sources with opposite phases at the frequency of 8 kHz, with the illusion effect schematically shown in the inset. The two virtual sources are set at (0, ±20) mm, respectively. (b) The phase discontinuity $\Delta\phi$ (black line) of metasurface and the amplitude of transmitted waves $|t|$ (dashed red line) of a two-turn unit cell with $l$ =15mm and $d$ =1mm, as a function of $h$ at 8 kHz. (c) The normalized amplitude of the measured $A_0$ wave field at the frequency of 8 kHz. (d) Same as (c), but at the frequency of 8.2 kHz. In subplots (a), (c) and (d), crests of the target pattern are plotted as gray lines.